\documentclass[sigconf,10pt]{acmart}

\AtBeginDocument{%
  \providecommand\BibTeX{{%
    \normalfont B\kern-0.5em{\scshape i\kern-0.25em b}\kern-0.8em\TeX}}}
\usepackage{caption}
\usepackage{subcaption}
\usepackage{mathtools}
\usepackage{multirow}
\usepackage{array}

\copyrightyear{2020} 
\acmYear{2020} 
\setcopyright{acmlicensed}
\acmConference[SIGMOD'20]{Proceedings of the 2020 ACM SIGMOD International Conference on Management of Data}{June 14--19, 2020}{Portland, OR, USA}
\acmBooktitle{Proceedings of the 2020 ACM SIGMOD International Conference on Management of Data (SIGMOD'20), June 14--19, 2020, Portland, OR, USA}
\acmPrice{15.00}
\acmDOI{10.1145/3318464.3386129}
\acmISBN{978-1-4503-6735-6/20/06}

\begin{document}
\fancyhead{}

\title{Taurus Database: How to be Fast, Available, and Frugal in the Cloud}

\settopmatter{authorsperrow=1}

\author{Alex Depoutovitch, \space Chong Chen,\space Jin Chen,\space Paul Larson,\space Shu Lin,\space Jack Ng, \space Wenlin Cui, Qiang Liu, Wei Huang, Yong Xiao, Yongjun He}
\affiliation{%
  \institution{Huawei Research Canada}
}
\renewcommand{\shortauthors}{Depoutovitch, et al.}

\begin{abstract}
Using cloud Database as a Service (DBaaS) offerings instead of on-premise deployments is increasingly common. Key advantages include improved availability and scalability at a lower cost than on-premise alternatives. In this paper, we describe the design of Taurus, a new multi-tenant cloud database system. Taurus separates the compute and storage layers in a similar manner to Amazon Aurora and Microsoft Socrates and provides similar benefits, such as read replica support, low network utilization, hardware sharing and scalability. However, the Taurus architecture has several unique advantages. Taurus offers novel replication and recovery algorithms providing better availability than existing approaches using the same or fewer replicas. Also, Taurus is highly optimized for performance, using no more than one network hop on critical paths and exclusively using append-only storage, delivering faster writes, reduced device wear, and constant-time snapshots. This paper describes Taurus and provides a detailed description and analysis of the storage node architecture, which has not been previously available from the published literature.
\end{abstract}



\maketitle
\section{Introduction}

As companies move their applications to the cloud, demand for cloud-based relational database  service (DBaaS) is growing rapidly. Amazon, Microsoft, Alibaba, and other cloud providers all offer such services. Most DBaaS offerings were initially based on traditional monolithic database software, essentially just running databases on (virtual) machines in the cloud, using either local storage or cloud storage.  Although simple to implement, this approach cannot provide what customers want from a cloud database service  \cite{Kossmann10}. From a customer's point of view, an ideal database service should be highly available, require no maintenance, and scale up and down automatically with database size and workload. It should also deliver high performance, be low cost, and users should pay only for resources actually used (pay-as you-go). These goals can't be achieved by running the same old database software in the cloud - the system must be redesigned from the ground up.

A good DBaaS architecture must simultaneously provide durability, scalability, performance, availability, and cost-effectiveness.  Having three persistent copies of data distributed across different hosts is usually considered sufficient to ensure durability \cite{Chansler12}. Scalability can be provided by distributing compute and storage resources across hosts. While distributing a database across multiple hosts improves durability and scalability, it can adversely affect availability. Using many hosts increases the probability of some hosts becoming unavailable, potentially resulting in the whole database becoming unavailable. Methods, such as quorum-based replication and eventual consistency, can increase availability, but both have drawbacks. Quorum-based replication may need a higher replication factor than what is required for pure data durability, increasing the cost. Eventual consistency is unacceptable for many existing applications.

In this paper, we present Taurus, a relational database designed specifically for cloud environments. Taurus builds on ideas from previous work and improves on them in several respects. Taurus separates compute and storage layers in a fashion similar to Aurora \cite{Verbitski17} and, like Socrates, separates the concepts of availability and durability \cite{Antonopoulos19}. Taurus offers similar benefits, such as read replica support, fast fail-over and recovery, hardware sharing, and scalability to 128TB.

The Taurus compute layer consists of a single master (primary) and multiple read-only replicas (secondaries). Data is divided into pages that are partitioned across multiple storage nodes. All update transactions are handled by the master. The master ships log records to the storage layer over the network. The storage layer writes them to reliable storage, stores database pages, applies received log records to bring pages up to date, and responds to page read requests. 

Taurus introduces several innovations, helping it achieve high availability and high performance at a low cost.  Our first contribution is a replication and recovery algorithm that achieves high availability with a replication factor no higher than what is required for durability and without sacrificing performance or strong consistency guarantees. This algorithm enables Taurus to achieve nearly 100\% availability for writes when considering uncorrelated storage failures. With only 3-way data duplication required for durability, Taurus achieves availability comparable to the 6-way quorum replication used by Aurora and better than the 3-way quorum replication used by POLARDB \cite{Cao18}.

A key observation is that data access patterns for the database log and database pages are significantly different. Database logs are used solely for data durability and are written and read sequentially. Log writes are frequent and performance is critical, but log reads are required only during generally infrequent failures, thus read performance is of lesser concern. As logs are necessary for durability, they need strong consistency guarantees. 

Unlike database logs, database pages are accessed randomly. Page read performance is critical, but write latency is less important because when a modified page needs to be written, the changes are already durable in the log. Database pages are versioned, and versioning can be used to provide strong consistency. Because of such differences, it is inefficient to use the same storage for logs and pages. Taurus separates log storage (Log Stores) from page storage (Page Stores). It uses different distribution and replication algorithms for log and page data. Each algorithm is optimized for its particular requirements.

Log reads are infrequent, and most log records are discarded without ever being read. A log record does not depend on other log records. Log records don't need to be written to specific Log Store servers, but rather can be written to any available Log Stores in a large pool, provided that the number of available Log Stores is enough for durability. In practice, a DBaaS deployment contains hundreds to thousands of nodes. A database can acknowledge writes as long as there are three available storage nodes, which for practical purposes should always be the case considering only uncorrelated failures.

In contrast, database pages are updated by applying log records and the previous version of a page is required to produce the next one. This requires a page to be assigned to specific Pages Store servers. Thus, the availability of a page depends on the availability of the specific Page Stores responsible for the page. However, in contrast to the traditional quorum-based replication algorithms, Taurus does not need a majority or even a plurality of Page Stores to be involved in each read or write operation. Strongly consistent log servers and page versioning can be used when deciding which Page Store replica is the most current one.  If the page version is known, Page Stores can be eventually consistent, which is known to improve service availability \cite{Fox97}. The lack of a quorum requirement allows Taurus to reduce the number of replicas to that required for durability.

Another key factor for a successful DBaaS is high performance. In distributed systems, performance significantly depends on the number of operations on critical paths that cross network boundaries. Our second contribution is a set of novel architectural choices that enable Taurus to achieve higher performance. Our tests show that Taurus can achieve up to 200\% higher throughput than MySQL 8.0 running with local storage. Taurus supports multiple read replicas and keeps the replica lag under 20ms even under high load. Most performance-critical operations in Taurus, such as writing logs and reading pages, require only one interaction between network-connected hosts. 

To avoid having the storage layer become a bottleneck, Taurus improves storage layer performance in two ways. First, the separation of Log Stores from Page Stores reduces the load on Page Stores. Second, the organization of Page Stores is optimized around a "the log is the database" model, never modifying data in place, and performing append-only writes. This approach has multiple advantages, improving write performance by 2-5 times, while simultaneously reducing device wear and consequently, the cost of the service. Also, append-only writes simplify consistency algorithms and snapshot generation. Finally, the separation of the storage layer into Log and Page Stores, allows Taurus read replicas to receive updates directly from Log Stores, bypassing the master and avoiding the master becoming a bottleneck. Taurus distributes resource consumption necessary to support read replicas evenly across the scalable storage layer.

The third contribution of this paper is a detailed description of the inner workings, performance optimizations, and design compromises in the storage layer. Although the available literature \cite{Antonopoulos19, Verbitski17} describes many optimizations done in the compute layer, it lacks details about storage  organization and the rationale for design decisions.

The remainder of this paper is organized as follows. In Section 2, we compare Taurus with available state of the art DBaaS implementations and discuss related work. Section 3 provides a high-level overview of the Taurus architecture.  Section 4 describes the details of the Taurus recovery algorithm and compares it with alternatives. Section 5 discusses recovery, and Section 6 covers read replica support. In Section 7, we do a deep dive into the organization of Page Stores. A performance evaluation is provided in Section 8, and we conclude in Section 9.

\section{Background and related work}
During the last few years, several relational database architectures designed for cloud environments have emerged: Amazon Aurora, Alibaba POLARDB, and Microsoft Socrates. \cite{Antonopoulos19, Cao18, Verbitski17}. Traditional databases deployed on cloud instances with private storage are also popular \cite{RDSMySQL, AzureSQL}. 

The main advantages of using traditional databases in the cloud include ease of implementation, no changes to and full compatibility with existing software. However, there are also disadvantages with this approach. With traditional databases, the database size is limited by the amount of locally attached storage. Network-attached storage can increase the maximum size of a database, but storage cost, network load, and update costs remain high and proportional to the number of replicas. This is because each replica needs to maintain its own copy of the database \cite{Verbitski17}. Adding a new read replica requires the entire database to be replicated, an expensive and time-consuming process proportional to the size of the database that significantly limits scalability. Size-of-data operations, such as taking a backup, also limit the database size; they simply take too long.

To address some of the above limitations, Cao et al. proposed running master and read replicas on top of a shared, distributed file system - PolarFS \cite{Cao18}. PolarFS provides a POSIX-like file interface, which makes it possible to keep the database code and architecture intact. Durability is provided by PolarFS through three-way replication using an optimized quorum protocol. Sharing storage between database replicas reduces storage cost and network load. Using a distributed file system enables storage scale-out and large database support. However, as the storage layer does not provide any database-specific processing, this architecture still inherits some limitations. These include write amplification, high network load due to page flushing, and limited performance and scalability as the master performs all processing.

Splitting the database system into compute and storage layers, and having each layer incorporate some of the database functionality solves many of these problems. Compute nodes ship only log records to the storage layer instead of full pages, and the storage layer knows how to update pages from log records. This approach, pioneered by Aurora, reduces network load and the load on the compute layer since the flushing of full pages is no longer necessary \cite{Verbitski17}. Aurora divides the database pages into 10GB shards distributed across shared storage nodes and uses quorum-based writes with six replicas for each shard. Using a quorum introduces availability problems as multiple replicas of a shard need to be online to ensure reads and writes are successful. For a quorum-based system with three nodes, two nodes must be online in order to serve reads and writes. However, the authors considered this inadequate for availability, so Aurora uses 6-node quorum, which improves availability.

Microsoft Socrates also relies on log shipping and page reconstruction at the storage nodes \cite{Antonopoulos19}. However, Socrates separates durability and availability by splitting the database into four layers: a compute layer that serves the same purpose as the corresponding layer in Aurora, a log layer that is specialized in fast log persistence and ensures log durability, a page server layer that applies log records to pages and serves page reads, and a storage layer that is used to ensure durability of the database data. Storing log records separately from pages improves database performance and allows it to use slower and cheaper storage for pages and faster, more expensive storage for log records.

Taurus also embraces the ideas of separating the database into different functional layers and separating the notions of availability and durability. Taurus takes these ideas further by using different replication and consistency techniques for logs and pages. This approach allows Taurus to simultaneously achieve higher availability, lower storage costs, and better performance. The Taurus replication algorithm provides higher availability for writes than the quorum replication used by POLARDB and Aurora, and uses only three data replicas to minimize storage costs. Taurus separates the database system into two physical tiers compared to Socrates' four-tier architecture. This results in lower network load and latency. In order to reduce read latency, Socrates caches all pages on local storage at the Page Server tier. In contrast, Taurus does not have intermediate tiers. It does not require caching as data can be quickly retrieved from storage devices with a single network hop.

In addition to the general database architectures described above, there are several cloud databases optimized for specific requirements. Spanner, a partially SQL-compliant database, was designed to handle geo-distributed transactions for read-intensive workloads and relies on two-phase commit and accurate clocks \cite{Corbett13}. Snowflake is an example of a cloud database optimized for large volume analytical data processing \cite{Dageville16}.  Vandiver et al. discuss experience of adapting the Vertica column store engine for cloud deployments \cite{Vandiver18}. 

Taurus builds on multiple previously developed techniques. Log structured storage was introduced by LFS \cite{Rosenblum92} and has been applied in several databases and key-value stores \cite{Ousterhout10,Vo12, Levandoski13, Sauer18}. Taurus Page Stores use this concept to store its persistent data. Eventual consistency and data versioning \cite{Lamport78} have been used by Dynamo DB \cite{DeCandia07} to achieve high availability at the cost of weaker consistency. Taurus uses a combination of consistency methods to achieve both availability and consistency. The gossip protocol for data replication was proposed by Demers et al. \cite{Demers87}, and has been used widely. Taurus uses a combination of gossip and central replication in order to overcome gossip limitations such as high network load.

\section{System Architecture}
Cloud environments are significantly different from traditional dedicated server environments. They violate implicit assumptions that were the basis for traditional database architectures. The first part of this section outlines why that this is the case, using MySQL as an example. This is followed by a high-level overview of the Taurus architecture emphasizing how it is optimized for running in a cloud environment.
\subsection{The cloud is different}

\begin{figure}[t]
\begin{center}
\includegraphics[width=0.9\linewidth]{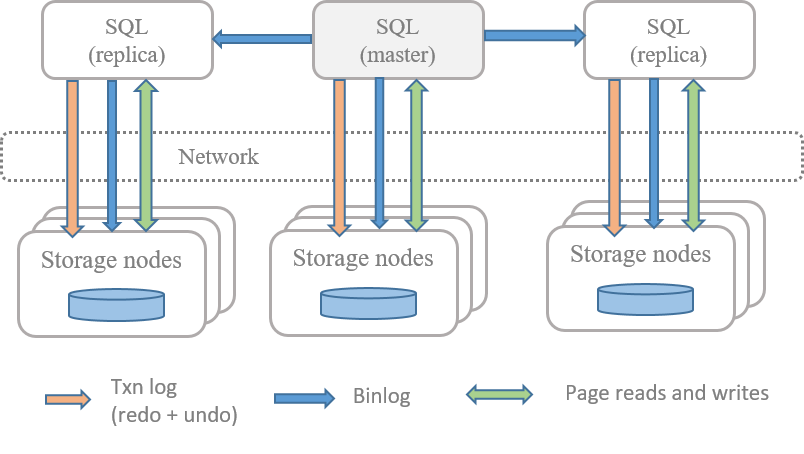}
\end{center}
\caption{MySQL with two replicas deployed in a cloud environment
\label{old-architecture}}
\end{figure}

Relational database systems have traditionally been designed assuming data is stored on dedicated local storage. If a database is required to be highly available, two or three copies of the database are typically maintained, a master and one or two replicas. The master handles both read and write requests. Each replica maintains a complete copy of the database and may also process read-only transactions. If the master fails or becomes unresponsive, one of the replicas takes over as the new master.

This architecture is well suited to on-premise deployments, but in a cloud environment, it wastes resources: network bandwidth, CPU cycles, memory space, storage space, and I/O bandwidth. Fig.~\ref{old-architecture} illustrates how data flows and is stored in a typical MySQL cloud deployment. Each MySQL instance runs on a separate VM, storing its data on virtual disks (volumes). The customer configures the VMs and virtual disks with fixed capacity and pays a fixed price, regardless of how much of this capacity is actually used. Virtual disks in the cloud typically store three copies of the data to guarantee high reliability and availability. As each of the three MySQL instances maintain its own copy of the database, this means that nine complete copies of the database are stored. This is clearly excessive and wasteful.

The arrows in Fig.~\ref{old-architecture} show how data flows. Replicas update their copy of the database by re-executing all update transactions received by the master. This means that every update transaction is executed three times, once per MySQL instance. Considering the replication done by storage, this means that every write is repeated nine times. 

In a cloud environment, this traditional architecture wastes resources, increasing the cost of the service. Scaling out compute by adding read replicas is slow and expensive because a completely new copy of the database needs to be created for each replica added. Large, multi-terabyte databases cannot be supported because operations, such as backup and restore, simply take too long.

\subsection{Taurus overview}

\begin{figure}[t]
\begin{center}
\includegraphics[width=0.9\linewidth]{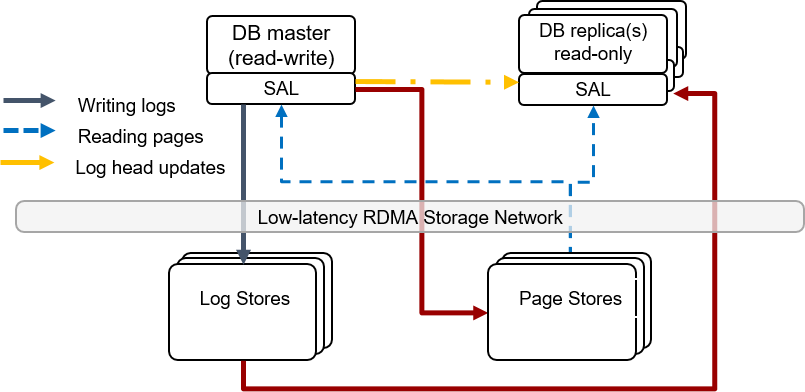}
\end{center}
\caption{Taurus components and layers}
\label{taurus-architecture}
\end{figure}

Taurus consists of four major logical parts: Log Stores, Page Stores, a Storage Abstraction Layer (SAL), and a database front end. These parts are distributed between two physical layers: a compute layer and a storage layer (Fig.~\ref{taurus-architecture}). Having only two physical layers minimizes the amount of data sent across the network and reduces the latency of requests, which can be completed with just one call over the network.

The database front end is a slightly modified version of MySQL, but PostgreSQL and other engines may be supported in the future. The front end is responsible for accepting incoming connections, optimizing and executing queries, managing transactions, and producing log records that describe modification made to database pages. The front end layer consists of one master replica that can serve both read and write queries and multiple read replicas that execute read queries only. In order to persist modifications to database pages, the log records must be made durable. 

The Log Store is a service executing in the storage layer responsible for storing log records durably. Once all log records belonging to a transaction have been made durable, transaction completion can be acknowledged to the client. Log Stores serve two purposes. First and foremost, they ensure the durability of log records. Second, they serve log records to read replicas so that the replicas can apply the log records to pages in their buffer pool. The master periodically communicates the location of the latest log records so that read replicas can read the latest log. The master also distributes log records to Page Store servers.

Page Store servers are also located in the storage layer. Taurus database is divided into small fixed-size (10GB) sets of pages called slices. Each Page Store server handles multiple slices from different databases and receives logs only for the pages that belong to the slices it is responsible for. A database can have multiple slices, and each slice is replicated to three Page Stores for durability and availability. The next sections describe the four major parts of Taurus in more detail.

\subsection{Log Store}
The primary function of the Log Store is to persist log records generated by the master and provide read access to them from any replica. The key storage abstraction provided by the underlying storage is called a PLog. A PLog is a limited-size, append-only storage object that is synchronously replicated across multiple Log Stores. The replication factor is determined by the desired level of durability, which in our implementation, is set to three.

Log Store servers are organized into a cluster. A typical cloud deployment has hundreds of Log Store servers. When the creation of a PLog is requested, the cluster manager chooses three Log Store servers to which the PLog will be replicated. It assigns a 24-byte identifier that uniquely identifies the PLog. Writes to a PLog are acknowledged only when all three Log Store replicas report a successful write. If one of the Log Stores fails to acknowledge the write within the expected time, the write is considered to have failed, no further writes are issued to this PLog, and a new PLog is created across another three Log Store Servers chosen by the cluster manager. This means that writes to Log Stores will always succeed as long as there are at least three healthy hosts available in the cluster. Writes are fast because if a Log Store is slow or a network packet is lost, writes are not retried to the old location, but rather sent to other less loaded or more reliable Log Stores.

Reads from the Log Store will succeed as long as there is at least one PLog replica available. Reads from PLogs occur in two cases. First, database read-only replicas read the log records recently written by the master replica. To accelerate such reads, Log Store caches recently written data in memory using a FIFO policy for eviction so that no disk access is required in most cases. Second, reads also occur during database recovery when committed log records must be read and sent to Page Stores.

The database log is stored in an ordered collection of PLogs, called data PLogs. The list of these PLogs is recorded in a separate metadata PLog and cached in memory on the database nodes. When a database is initialized, a metadata PLog and PLogs storing the database log are automatically created. Writes to the metadata PLogs are governed by the same rules as data PLogs. When a new data PLog is created or removed, all metadata is written in one atomic write to the metadata PLog. When a metadata PLog reaches its size limit, a new metadata PLog is created, the latest metadata is written there, and the old metadata PLog is deleted.

\subsection{Page Store}
The main function of Page Stores is to serve page read requests coming from the database master server or read replicas. A Page Store must be able to recreate any version of a page that a database front end may request so a Page Store must have access to all log records for the pages that it is responsible for. This requirement prevents us from switching Page Stores in same way as we switch Log Stores when they become unavailable and makes achieving high availability more challenging.

When the master modifies a page, it assigns the page a version, a monotonically increasing logical sequence number (LSN) that uniquely identifies and establishes order among all changes to a database. Each page version is identified by its page ID and LSN.

The SAL communicates with a Page Store through an API that exposes four major methods:
\begin{enumerate}
\item \textit{WriteLogs} is used to ship a buffer with log records
\item \textit{ReadPage} is used to read a specific version of a page
\item \textit{SetRecycleLSN} is used to specify the oldest LSN of pages belonging to the same database that the front end might request (recycle LSN)
\item \textit{GetPersistentLSN} returns the highest LSN that the Page Store can serve
\end{enumerate}

A Page Store is responsible for multiple slices from different databases, and each slice is identified by a unique identifier that is passed to each of the above methods. Starting from the first write to a page, each change to a page is received as a log record passed to \textit{WriteLogs}. A Page Store continuously applies incoming log record in the background to generate and store new versions of pages.

Whenever an SQL front end needs to read a page, SAL calls \textit{ReadPage} specifying the slice, id of the page, and the version of the page that it needs. A Page Store must be able to serve older versions of a page because a read replica may lag behind the master in its view of the database. In order to get a consistent physical view of the database, the replica specifies the LSN up to which the Page Store must have all log records in order to serve the read request. This way the Page Store makes sure that it returns a page version that is up to date and is not ahead of what the read replica expects.

Because it takes resources (memory and disk space) to store multiple versions of a page, the SQL layer must periodically call \textit{SetRecycleLSN} to communicate to Page Stores the oldest LSN it may request. Finally, because data replicated between Page Store replicas using an eventually consistent model, the master needs to know the latest version of the pages each Page Store replica has knowledge of. This information is retrieved using the \textit{GetPersistentLSN} API call. We describe the Page Store internals and the replication of data between Page Store replicas in more detail later.

\subsection{Storage Abstraction Layer} \label{sal-section}
The Storage Abstraction Layer (SAL) is a library linked to the database server that isolates the existing database front end (such as MySQL or PostgreSQL) from the underlying complexity of remote storage, slicing of the database, recovery, and read replica synchronization. SAL is responsible for writing log records to Log Stores and Page Stores and reading pages from Page Stores. SAL is also responsible for creating, managing, and destroying slices in Page Stores and mapping pages to the slices.
Whenever a database is created or expanded, SAL selects Page Stores and creates slices on the selected Page Stores. Whenever the master decides to flush log records, the log records are sent to SAL. To avoid small writes, log records are accumulated and flushed to Log Stores as a group called a database log buffer. SAL first writes the database log buffer to the currently active Log Store replicas to guarantee their durability. Once the log records are successfully written to all Log Store replicas, the write is acknowledged to the master, and the log records are distributed to the per-slice buffers. The slice buffers are flushed either when they become full or after a timeout.

An important value that SAL maintains is called the cluster visible (CV) LSN. The CV-LSN represents an LSN, global to the database, at which all pages of the database are internally consistent with each other (such as internal pages of a B-tree). For example, if a B-tree contains a page B, which is a child of a page A, and if page B is split into page B and C, the operation has to be atomic. In this case, the CV-LSN of the database will jump from the previous value to the value that is the largest of the new LSNs of pages A, B, and C. Further, the CV-LSN is always set to the latest point up to which redo log records have been persisted in all slices (but not necessarily all slice replicas) without any gaps. Thus, at least one Page Store can serve a page with LSN equal to or larger than CV-LSN. With the CV-LSN, SAL can establish a consistent and forward-moving state of the database across all logical components in Taurus (master, read replicas, and Page Stores). The CV-LSN is advanced in increments of the last LSN of each database log buffer. A database log buffer may contain records targeting one or more slices. SAL advances CV-LSN only when both of the following conditions are met:

\begin{enumerate}
\item The database log buffer has been successfully written to the Log Stores
\item All per-slice buffers that contain records from this log have been successfully written to at least one of the Page Store for each slice containing pages matching the log records of the group flush
\end{enumerate}

Per-slice buffers are submitted for flushing only after the database log buffer has been written to the Log Stores. Consequently, Page Stores can never contain log records that have not been written to Log Stores. Each per-slice buffer contains only a subset of records from the database log buffer, as the database buffer has records for all slices. Thus, per-slice buffers are flushed less frequently and may contain log records that correspond to multiple database log buffers. To advance the CV-LSN according to the rules above, SAL has to track the many-to-many relationship between the database log buffers and the per-slice buffers.

\subsection{Database front end}
Currently, Taurus uses a slightly modified version of MySQL 8.0 as a database front end. Modifications include forwarding log writes and page reads to the SAL layer and disabling buffer pool flushing and redo recovery for the master node. Modifications to read replicas include updating pages in the buffer pool using records read by SAL from Log Stores and a mechanism for assigning read views to transactions.

\section{Replication}
Data durability and service availability are critical characteristics of a database service. In practice, having three copies of data is generally considered sufficient for durability \cite{Chansler12}. However, three copies may not be enough to reach the required availability level. Maintaining three consistent copies means that all of them must be online for the system to be available. When a database is sharded across many nodes, the probability of some node being unavailable grows exponentially with number of nodes. 
In Taurus, we solve this availability problem using a novel approach of having different replication policies for log and data, coupled with a novel recovery algorithm. We first describe the write path, then the read path, and finally present theoretical availability estimates and compare them with other replication strategies. 

\subsection{Write path}
\begin{figure}[t]
\begin{center}
\includegraphics[width=1.0\linewidth]{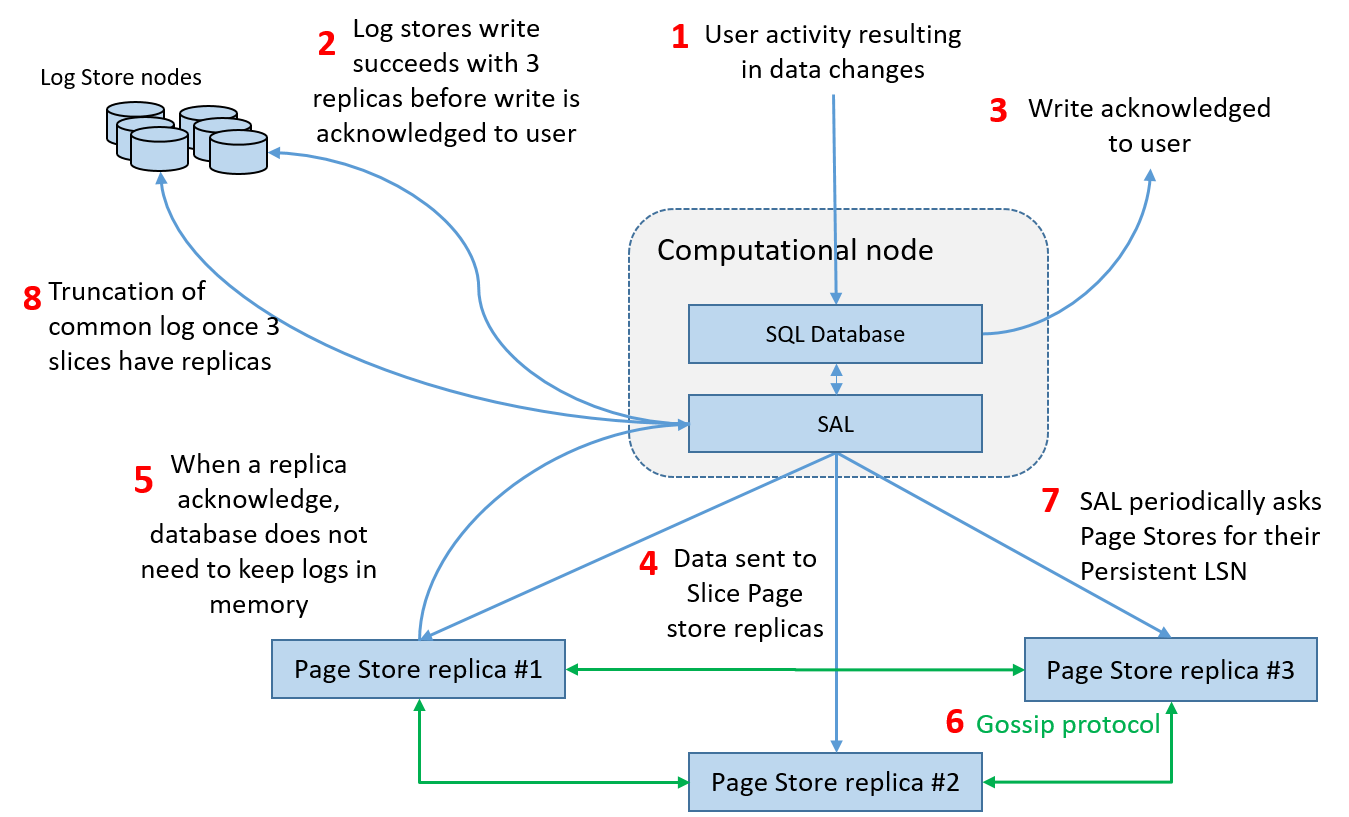}
\end{center}
\caption{Taurus write path}
\label{write-path}
\end{figure}

The write flow is summarized in Fig. \ref{write-path}. First, user transactions result in changes to database pages, which generates log records describing the changes. To make log records durable, SAL writes them to PLogs located on the three available Log Stores (Step 2). To avoid fragmentation but balance the load across multiple Log Stores, SAL limits PLog sizes to 64MB after which it is sealed and a new PLog is created on some Log Store in the cluster, taking into account the free space and load on the different Log Stores. 

Once all Log Stores acknowledge the write, there are three durable copies of the data. The database then considers the data to be persistent and the write complete. Transactions whose commit depends on the write can be marked as committed (Step 3).  The important point is that the database is available for writes as long as three or more Log Stores are available in the cluster. For clusters of thousands of nodes, this means practically 100\% write availability (considering only uncorrelated failures).

Once a log record for a page has been written to the Log Stores, SAL copies it into the write buffer of the slice that is responsible for the page. When a slice write buffer is full, or after a timeout, the buffer is sent to the Page Stores hosting the slice (Step 4). Each buffer contains a slice ID and sequence number to allow Page Stores to detect missing buffers. SAL waits for a reply from one of the Page Stores after which the buffer is released and can be reused (Step 5). The largest LSN that has been sent to a slice is called the slice flush LSN. Page Stores that host replicas of the same slice also periodically exchange messages with each other using a gossip protocol to detect and recover missing buffers (Step 6). Steps 7 and 8 are described in the Log Truncation section.

While writes to Log Stores are complete only when all three replicas have replied, SAL only waits for a single Page Store to confirm a successful write. This has several advantages. First, the probability of a successful write is much higher, as it requires only one node out of three to be available. This ensures high availability, even in the case of failures or temporary glitches common when a large number of nodes are involved. Note that data durability is not compromised because log records have already been persisted on Log Store nodes. Second, write latency is minimized because it depends only on the fastest node to reply rather than the slowest. Third, SAL is no longer responsible for ensuring that every log record reaches all corresponding Page Stores. Consequently, SAL keeps log records in memory for a shorter time consuming less CPU and network bandwidth re-sending log records. Re-sending log records as needed is offloaded to the Page Stores themselves. This is important for scalability because there is only a single master node  but many Page Stores nodes that can share the load.

\subsection{Read path}
Database front ends read data at page-level granularity. When reading or modifying data, the corresponding page must be present in the buffer pool. When the buffer pool is full, and we need to bring in a page, some page has to be evicted. We modified the eviction algorithm so that a dirty page cannot be evicted until all of its log records have been written to at least one Page Store replica. Thus, until the latest log record reaches a Page Store, the corresponding page is guaranteed to be available from the buffer pool. After that, the page can be read from a Page Store. 

For each slice, SAL maintains the LSN of the last log record sent to the slice. Whenever the master node reads a page, the read goes to the SAL, which issues a read request accompanied by the above LSN. Reads are routed to Page Stores that are known to respond with the lowest latency. If the chosen Page Store is unavailable or did not receive all log records up to the provided LSN, an error is returned, and SAL tries the next Page Store hosting the desired page, iterating through the replicas until it finds one that can execute the request.

\begin{table*}[t]
\setlength\extrarowheight{2pt}
\centering
\begin{tabular}{|c|c|c|c|c|c|c|c|c|}
\hline
\multirow{2}{*}{Replication method} &\multicolumn{2}{|c|} {Probability of non-availability}& \multicolumn{2}{|c|} {x = 0.15} & \multicolumn{2}{|c|} {x = 0.05} & \multicolumn{2}{|c|} {x = 0.01} \\
&\hspace*{20pt}Write\hspace*{20pt}& Read & Write & Read & Write & Read & Write & Read \\ \hline
$N = 6, N_W = 4, N_R = 3$ & $20 * x^3$ & $15 * x^4$&$7*10^{-2}$ &$8*10^{-3}$ &$3*10^{-3}$ &$10^{-4}$ &$2*10^{-5}$& $2*10^{-7}$\\ \hline
$N = 3, N_W = 2, N_R = 2$ & $3 * x^2$ & $3 * x^2$&$7*10^{-2}$ &$7*10^{-2}$ &$8*10^{-8}$ &$8*10^{-3}$ &$3*10^{-4}$ &$3*10^{-4}$ \\ \hline
$N = 3, N_W = 3, N_R = 1$ & $3 * x$ & $x^3$&$5*10^{-1}$ &$3*10^{-3}$ &$2*10^{-1}$ &$10^{-4}$ &$3*10^{-2}$ &$10^{-6}$ \\ \hline
Taurus & 0 & $x^3$&0 &$3*10^{-3}$ &0 &$10^{-4}$ &0 &$10^{-6}$ \\
\hline
\end{tabular}
\caption{Comparing the probability of the storage being unavailable for Taurus and common quorum replication variants}
\label{availability-theory}
\end{table*}

\subsection{Log truncation}
Log Stores receive a continuous stream of log records. Unless these log records are eventually deleted, Taurus will run out of space as the database continues to be updated. A log record cannot be discarded from Log Stores until the record has been successfully written to all slice replicas and has been seen by all database read replicas. It would be prohibitively expensive to track the persistence of each log record independently, therefore we track persistence based on LSN. For each of its slices, a Page Store tracks a slice persistent LSN, which is the LSN up to which the Page Store has received all log records for the slice. This LSN can be communicated back to SAL either explicitly by a call of the \textit{GetPersistentLSN} method or implicitly by piggybacking the current value of the persistent LSN on the response of \textit{WriteLogs} or \textit{ReadPage} calls. This reduces traffic for databases with a large number of frequently updated slices - see Fig. \ref{write-path} Step 7. 

For each replica of each slice, SAL tracks the persistent LSN of the corresponding Page Store. The minimum of persistent LSNs across the slices that have log records that have not reached all slice replicas is called the database persistent LSN. SAL periodically saves this value for recovery purposes. SAL also keeps track of the LSN range for each PLog containing log records. If all records in a PLog have an LSN smaller than the database persistent LSN, this PLog can be deleted (Step 8), thereby truncating the log. This way, we guarantee that each log record is always replicated on at least three nodes.

\subsection{Comparison with quorum replication}

The most widely used strongly consistent replication techniques are based on quorum replication \cite{Gifford79}. When an item is replicated across N nodes, each read or write must receive replies from N\textsubscript{R} nodes for reads and N\textsubscript{W} nodes for writes. To ensure strong consistency the condition N\textsubscript{R} + N\textsubscript{W} > N must be satisfied. Many existing systems use different variants of quorum replication. For example, synchronous replication used in RAID 1 disk arrays often uses N = 3, N\textsubscript{R} = 1, N\textsubscript{W} = 3, Polar DB uses N = 3, N\textsubscript{R} = 2, N\textsubscript{W} = 2, while Aurora uses N = 6, N\textsubscript{R} = 3, N\textsubscript{W} = 4. 

For the discussion below, we consider failures of each replica to be independent ignoring the correlated failures that affect multiple replicas at the same time, such as global power outages. Let's assume that the probability of a single node being unavailable is x. A write fails when between N - N\textsubscript{W} + 1 and N nodes are unavailable at the same time. If we sum up all combinations of nodes being unavailable, the probability of not being able to complete a write is calculated as follows:
\begin{equation}
P_w = \sum\limits_{i = N - N_W + 1}^{N} C_N^{i} * x^{i}(1 - x)^{N - i}
\label{p_w}
\end{equation}
Similarly, for reads, the formula is:
\begin{equation}
P_r = \sum\limits_{i = N - N_R + 1}^{N} C_N^{i} * x^{i}(1 - x)^{N - i}
\label{p_r}
\end{equation}

Unlike pure quorum writes, Taurus log writes don't need to land on specific Log Store nodes, so formula \ref{p_w} is not applicable. The probability of the storage layer being unavailable for writes due to independent node failures is close to zero for a cluster of hundreds of nodes because if a chosen node is unavailable, any other node can be chosen instead. Individual node failures affect latency, as failed writes have to be retried with a different set of Log Store nodes, but they don't affect availability.

For reads, every Page Store node can decide based on its persistent LSN whether it can service a read, and if not, it returns an error instructing the SAL to try another node. In the very rare case where no node can serve a read due to cascading failures, SAL recognizes this situation and repairs data using Log Stores (see the Page Store Recovery section). In this case, there is a performance penalty, but the storage layer continues to be available. The only situation that SAL cannot recover from is when all Page Stores that contain a slice replica are unavailable. The probability of this is $x^3$.

In Table \ref{availability-theory}, we assume that x $\ll$ 1, take into account only terms with the lowest exponent index, and obtain approximate formulas for the probability of storage being unavailable using equations \ref{p_w} and \ref{p_r}. We also add three example values for x to compare actual probabilities of storage being unavailable. Taurus is always available for writes in case of uncorrelated failures. For reads, Taurus provides the same or better availability than  quorum replication for all configurations and values of x except when x = 0.01, and the number of nodes in the quorum is 6. However, in this case, quorum replication has to use twice as many nodes compared to Taurus replication, and the probability of not being able to read data is low in both cases ($2*10^{-7}$ and $10^{-6}$ respectively).

\section{Recovery}

\iftrue
\begin{figure*}
\centering
\begin{subfigure}{0.52\linewidth}
  \centering
  \includegraphics[width=1.0\linewidth]{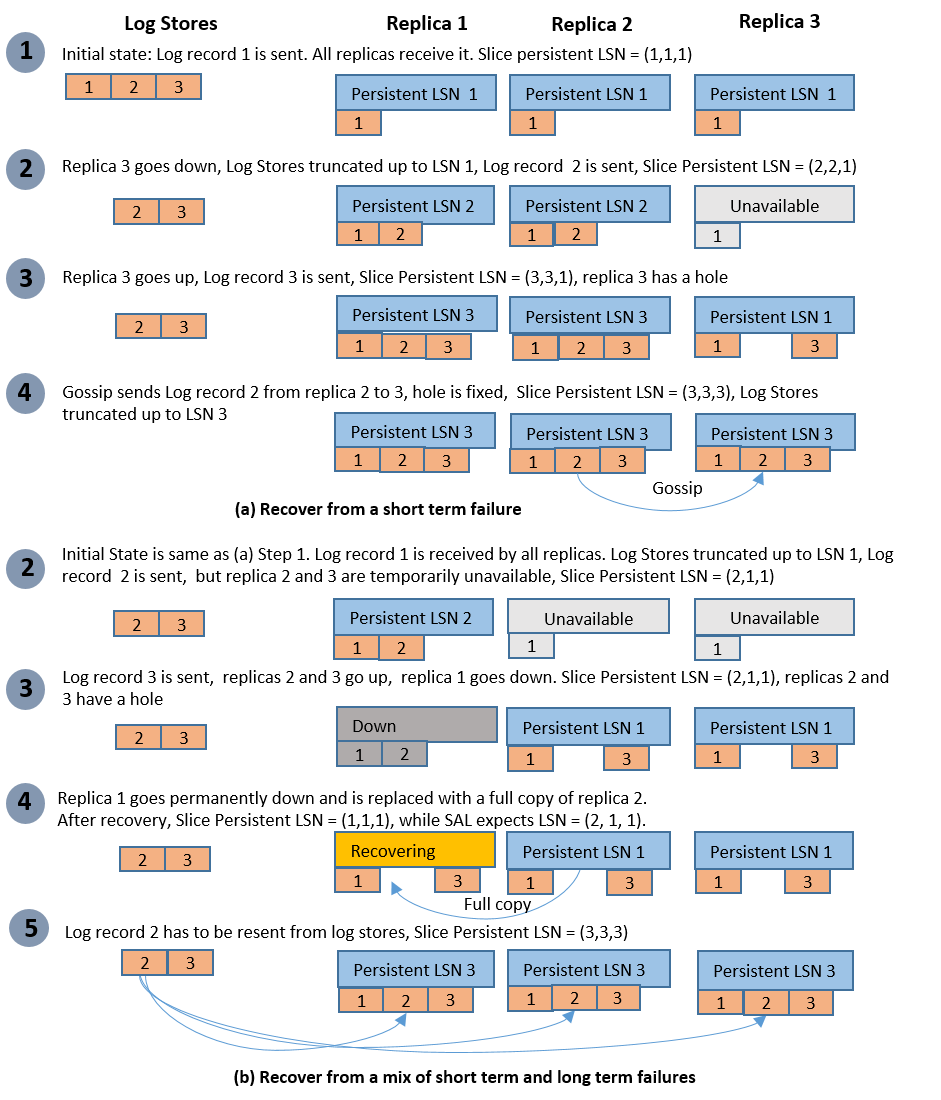}
  \label{pg-recovery-a}
\end{subfigure}%
\begin{subfigure}{0.52\linewidth}
  \centering
  \includegraphics[width=1.0\linewidth]{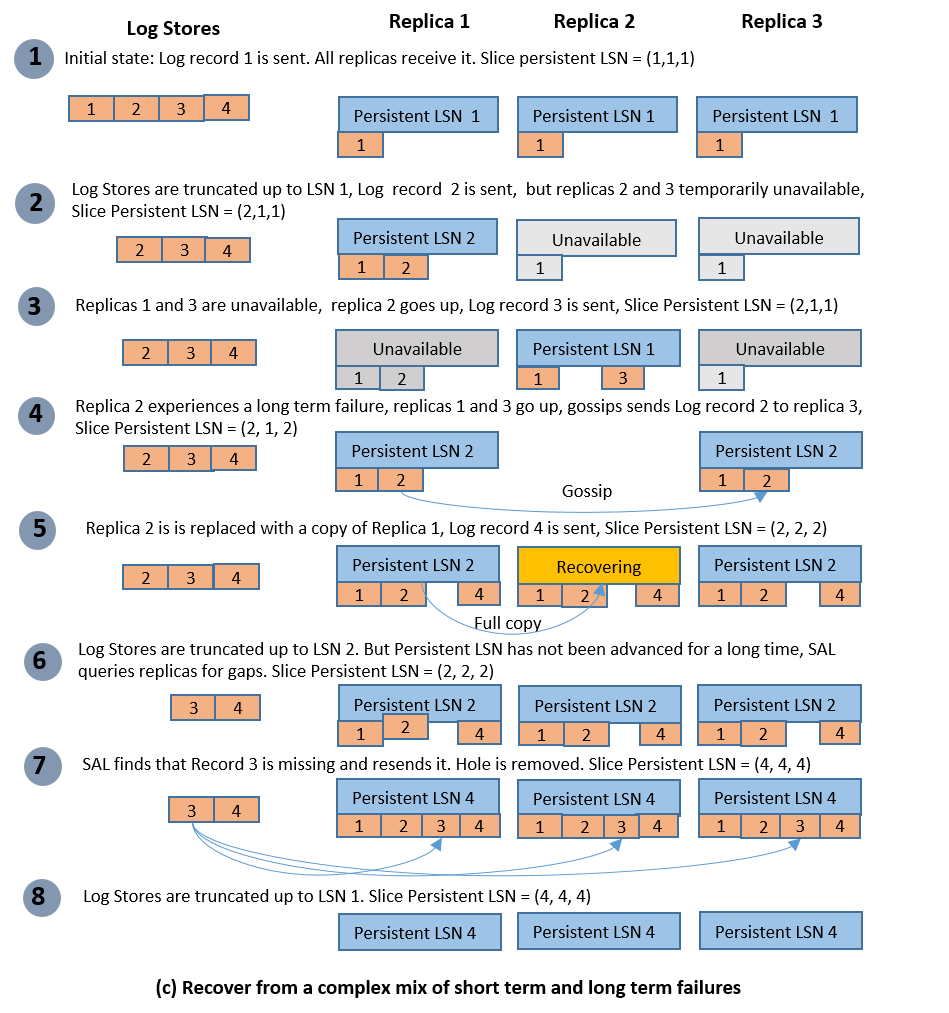}
  \label{pg-recovery-b}
\end{subfigure}
  \caption{Page Store recovery}
\label{pg-recovery}
\end{figure*}
\fi

Four types of nodes comprise a database instance: nodes that run master and read replica front ends, nodes that run Log Stores, and nodes that run Page Stores. Any combination of nodes may fail at any time. For a large database distributed across many Page and Log stores, individual node failures are expected to be a routine event. There are multiple types of failures: hardware, software, and network, to name a few, but for our purposes we define a node failure as an event during which the node does not serve incoming requests within a specified time limit.

An important design objective for recovery is to make failures and subsequent recovery invisible to user applications to the greatest extent possible. In our design, failures of Log Stores and Page Stores are invisible to applications; they will only be aware of front end failures. 

Storage nodes containing Page and Log Stores are constantly monitored by a recovery service. If a failure is detected, it is initially classified as a short term failure, and the corresponding node continues to be monitored. If the node remains unavailable for a longer period of time (tens of minutes), the failure is classified as a long term failure. The maximum length (currently, 15 minutes) of the short term failure is set sufficiently small that having only two available replicas of data does not violate durability guarantees.

\subsection{Log Store recovery}
Log Store failures are easy to handle and recover from. As described earlier, as soon as a Log Store becomes unavailable, all PLogs located on the Log Store stop accepting new writes and become read-only. Thus, no recovery is needed after a short term failure. When a long term failure is diagnosed, the failed node is removed from the cluster and PLog replicas from the failed node are recreated on the remaining cluster nodes from the available replicas.

\subsection{Page Store recovery}

Recovering from Page Store failures is more complicated. When a Page Store comes back online after a short term failure, it initiates the gossip protocol with other Page Stores that maintain replicas of the slices that this Page Store hosts. The gossip protocol recovers log records that a Page Store has missed. An example of the recovery process is illustrated at Fig.~\ref{pg-recovery}(a). For simplicity, we uses LSN 1, 2, 3 as LSN of log records.  There, log record 2 is copied from replica 2 by the gossip protocol after slice replica 3 is back online at step 4.

When a long term failure is detected, the cluster manager removes the failed node from the cluster and redistributes slice replicas that have been stored on the failed Page Store node among the remaining Page Store nodes. A recovering slice replica is initially empty. It can immediately begin accepting \textit{WriteLogs} requests, but because it does not have the necessary older log records, it cannot serve read requests. Next, the recovering slice requests the latest versions of all pages from one of the Page Stores that has a replica of the slice. Once all pages have been received, the slice replica can serve both reads and writes.

The above two scenarios are the most common but, until a log record is successfully processed by three Page Stores, there is a chance of it being lost due to Page Store failures. This can happen when multiple slice replicas fail intermittently within a short time, and Page Stores that have received a log record experience a long-term failure. 
\iftrue
An example of this situation is illustrated by Fig.~\ref{pg-recovery}(b). At Step 2, replicas 2 and 3 are offline for a short time. Log record 2 is acknowledged by replica 1 and dismissed by SAL. At step 3, replica 1 goes permanently offline due to a long term failure before the gossip protocol copies the missing record to replicas 2 or 3. At step 4, replica 1 will be restored to be a copy of either replica 2 or 3, missing the log record 2. In this case, no Page Store will contain record 2, and the record cannot be recovered by the gossip protocol. However, because not all slice replicas have acknowledged reception of log record 2, its copy still remains in the Log Stores. As described previously, the SAL regularly requests a persistent LSN from all replicas of slices that have been updated recently. In this scenario, the persistent LSN reported by replica 1 will be reduced from 2 to 1. When SAL detects that the persistent LSN reported by a slice replica decreases, it reads all log records from the Log Stores beginning from the smallest persistent LSN reported by the slice replicas and resends them to the corresponding Page Stores.

Detecting missing records by checking for decreases of persistent LSNs is fast but not sufficient. It works when a Page Store has received a complete sequence of log records (no holes), so it can advance the slice persistent LSN. However, if there are holes, a Page Store cannot advance the slice persistent LSN when a new record arrives even though the log record has been acknowledged and dismissed by the SAL. To address this scenario, SAL periodically retrieves the persistent LSN from all slice replicas and compares it with the slice flush LSN. If SAL detects that the persistent LSN does not advance forward and is smaller than the flush LSN, SAL requests the list of LSN ranges that have not been received by each slice replica. If SAL detects that some log records are missing from all Page Stores, it reads the missing records from the Log Stores and resends them to Page Stores. Fig.~\ref{pg-recovery}(c) illustrates an example of this scenario. At step 2, replica 2 and 3 go down; at step 3, replica 1 goes down, and replica 3 goes up; at steps 4 and 5 replica 2 experiences a long term failure and is replaced. At step 6, all replicas are up, but record 3 is missed from all of them and cannot be recovered by the gossip protocol. Unlike the Fig.~\ref{pg-recovery}(b) scenario, the slice persistent LSN doesn't decrease. SAL detects it using the method described above and resends record 3 at step 7.
\fi

Taurus is designed to support thousands of Page Stores, and the gossip protocol is known to be costly when invoked frequently with a large number of hosts \cite{Birman07}. For this reason, gossip is invoked automatically only every 30 minutes for each slice. To minimize the effect on availability during such an extended period, we rely on SAL. SAL monitors all slices that it has sent log records to. If a replica of such a slice has not advanced its persistent LSN accordingly, it means that some fragments are missing. If SAL detects that missing log fragments have not been recovered after a short time, it triggers the gossip protocol for this specific slice, to accelerate recovery of the missing fragment. If SAL detects that a log fragment is missing from all slice replicas, it rereads the log records from Log Stores and resends them to the Page Stores.

\subsection{SAL and database recovery}
Because SAL is a library, SAL and the database front end fail and recover together whenever the database process restarts. This can happen when there is an unrecoverable software failure, a restart, or when a cluster manager detects that the master is unavailable and spawns a new master or promotes a read replica to take over the master role (see the "Read Replicas" section).  

The database recovery process involves two main steps: 1) SAL recovery and 2) database front end recovery. SAL recovery happens first, and its primary goal is to ensure that all Page Stores that contain slices from the database have all the log records that were persisted in the Log Stores before a crash. SAL reads the last saved value of the database persistent LSN and uses it as a starting point from which to begin reading the log. Only log records that are missing from all slice replicas are sent again to the corresponding slices. Some log records may be resent even though some Page Stores contain them, but this is safe as Page Stores disregard log records that they have already received. This step is equivalent to the redo phase in traditional database recovery. After SAL recovery is complete, the database can accept new requests. In parallel with accepting new requests, the database front end performs the undo stage by rolling back changes made by transactions that were uncommitted at the time of the crash. Redo recovery must be completed before accepting new transactions because redo makes sure that the Page Store can serve reads of the latest versions of pages. Undo recovery relies on undo records stored in dedicated rollback pages. Thus, SAL must guarantee that all rollback pages are up-to-date in the Page Stores before undo processing in the SQL layer can begin.

\section{Read replicas} \label{read-replicas-section}
Read replicas  allow fast fail-over and scale-out capability for read workloads. Separate storage and compute layers allow each read replica to have direct access to the same storage as the master, and updates made on the master are automatically visible to read replicas.


\begin{figure}[t]
\begin{center}
\includegraphics[width=1.0\linewidth]{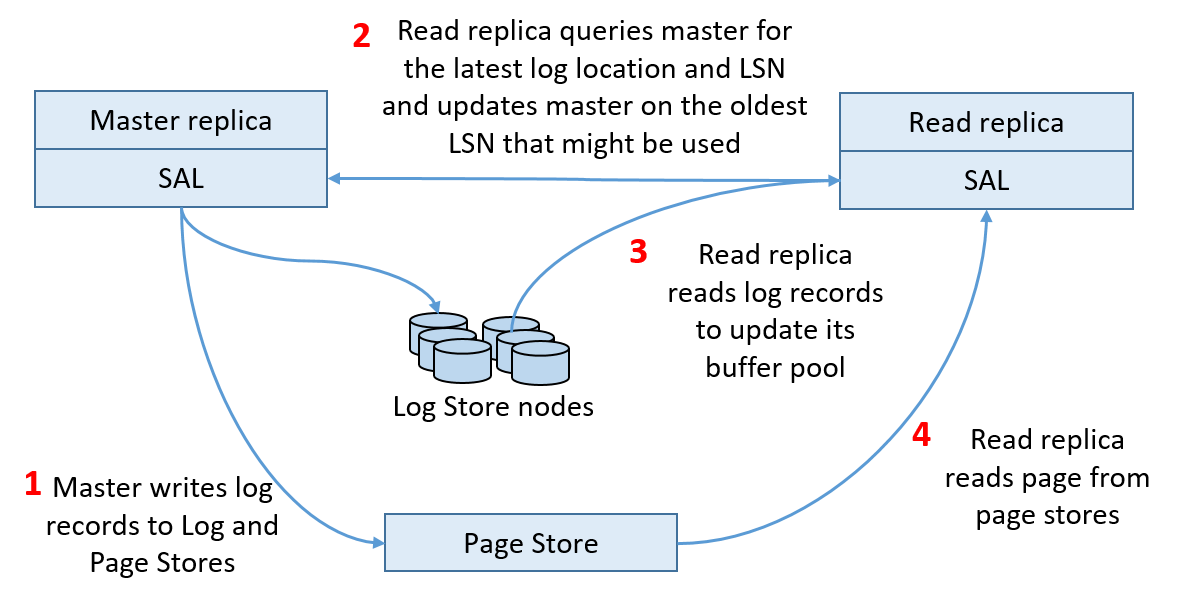}
\end{center}
\caption{Read replica workflow}
\label{read-replicas}
\end{figure}

Fig. \ref{read-replicas} shows how a read replica is kept up-to-date. When the master updates the database (step 1), a read replica gets messages from the master with the location of log records in Log Stores, changes to the list of slices, and finally the LSN of the last database change (step 2). Next, the replica reads all log records from the Log Stores in order to update pages in its buffer pool (step 3). Log Stores maintain a FIFO write-through cache, expecting that log records recently written will be read very shortly by read replicas. This cache nearly eliminates read I/O to the Log Stores. Finally, read replicas also read pages from the Page Stores (step 4) as needed.

An alternative design would be to stream all log records directly from the master to every read replica. However, this approach would result in the master becoming a bottleneck. Not only would the master need to spend CPU and memory transmitting log records, but its network interface may become a bottleneck. With write-intensive workloads generating 100MB/s of log records and 15 read replicas, the master would need to send over 12 Gbps of data just to read replicas.

Our solution is to make the master transfer only the location of the data in the Log Stores and leave it to each read replica to read the data. The read replica will need to receive messages from the master to know when and where new log records have been written, how the pages are distributed across Page Stores, and the Page Store's persistent LSNs. The master delivers messages that describe incremental changes of the above information to read replicas. Each message includes a sequence number so that missing messages can be detected. In this case, a read replica requests full data as when a new replica is registered.

Data is shared, and it can be modified by the master node without synchronizing with read replicas. Thus an important challenge is how read replicas can maintain a consistent view of the data. There are two types of consistency that we need to be concerned with. First, physical consistency refers to the consistency of internal structures in the database, such as b-tree pages. For example, when a thread is splitting a page in an index tree, changes involve multiple pages, and another thread that traverses the same tree must observe changes to pages involved as if they are done atomically. On the master, page consistency is achieved by locking the pages when they are being modified. However, it would be prohibitively expensive to coordinate locks with read replicas. To avoid explicit synchronization, the master writes log records in groups, always setting the group boundary at a consistent point.  Read replicas read and apply log records atomically per these group boundaries. The LSN of the last record processed by a read replica represents the replica's physical view of the database and is called the replica visible LSN. The replica visible LSN is always set at a group boundary keeping its database view physically consistent. 

A read replica reads and parses the log from Log Stores, recognizes log record group boundaries, and continuously advances its visible LSN. The read replica takes care not to advance its visible LSN ahead of the slice persistent LSN obtained from the master. This way it avoids a situation where a Page Store might not be able to serve the read replica's read page requests.  When a read transaction needs to access pages, it creates its own physical view of the database by recording the current replica visible LSN, called a transaction visible LSN (TV-LSN). Different transactions can have different TV-LSNs. While the read replica keeps advancing its visible LSN,  a transaction's visible LSN can lag behind. The read replica keeps track of the smallest TV-LSN and sends it to the master. The master collects the LSNs, chooses the minimum, and sends it to Page Stores as a new recycle LSN. Page Stores must ensure that they can serve any page version created after a recycle LSN. When a read transaction finishes an operation that requires physical consistency (e.g., an index lookup),  it releases its TV-LSN.  Since such operations are usually short, read replicas advance their recycle LSNs fairly quickly permitting the Page Store to purge old versions reducing the storage footprint.  

Many databases, including MySQL, maintain multiple versions of rows to reduce conflicts between readers and writers. Logical consistency refers to the consistency of user data as required by the transaction's isolation level. When a write transaction commits on the master, a commit record is written to the log. After parsing the log,  the read replica can update its active transaction list. When a read transaction starts on the read replica, it records the active and committed transaction list. This list determines a transaction's logical data view, i.e., which data is visible to a transaction.

The buffer pool on a read replica can store multiple versions of the same page. As the read replica reads and parses the log, it applies the log records to the buffer pool pages and produces newer versions of the pages. This way, the read replica already has most of the frequently used pages in its buffer pool, thus relieving pressure on Page Stores.

\section{Page Store design}

\begin{figure}[t]
\begin{center}
\includegraphics[width=1.0\linewidth]{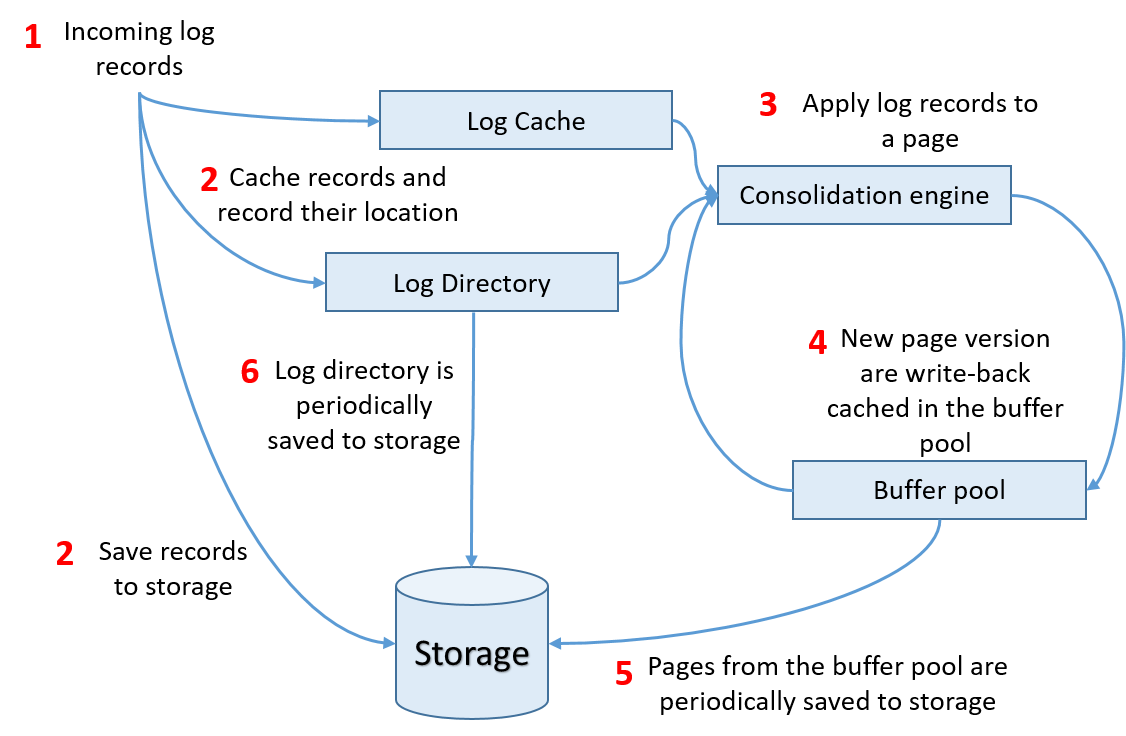}
\end{center}
\caption{Page Store major components and workflow}
\label{Page Store}
\end{figure}
The main function of Page Stores is to serve page read requests from the master and read replicas. Page modifications arrive as a stream of log records. For write-intensive workloads, Log records can arrive at a rate of a few million entries per second. A Page Store must be able to  apply log records and produce and persist new versions of pages at this rate, a process that we call log consolidation. For reliability, each database slice must be replicated across several Page Stores. 

These requirements guided several design decisions in Taurus. First, each Page Store performs log consolidation independently of other Page Stores as synchronizing consolidation among replicas of the same slice is prohibitively expensive.  Second, disk writes are append-only as append-only writes are 2-5 times faster than random writes and cause less device wear for flash-based storage \cite{Changman15}. Third, data needed for consolidation, i.e. base pages and log records, must be cached in memory as there may be thousands of log records fragmented on disk and storage cannot sustain reading log records at the required rate. Page Stores are based on a "the log is the database" paradigm \cite{Vo12}. This raises the  issue of how to quickly locate all necessary log records to produce a requested version of a page. For each slice, there is a data structure called the Log Directory. It keeps track of the location of all log records and the versions of the pages hosted by the slice,  i.e., information needed to produce pages. The Log Directory is implemented as a lock-free hash table \cite{Michael02}  where the keys are page IDs. Because hot pages are cached by database front ends, and writes are acknowledged as soon as they are persisted by Log Stores, the Page Store is not usually on the critical transaction path. However, if the consolidation is unable to keep up with the incoming log stream for some time, the Log Directory may grow large. To prevent unlimited growth, the SAL throttles log writes on the master.

Figure \ref{Page Store} illustrates the major components of the Page Store, and their interaction is described below. Logs records are received from the SAL in ordered groups, called log fragments (step 1), as described in the Section \ref{sal-section}. Log fragments are immediately appended to the slice's log on disk (step 2), cached in memory in the log cache, and the location each log record is added to the log directory (step 3). Log caching is extremely important because reading log records one by one during consolidation would be too slow. Consolidation accesses the Log Directory to locate an existing page and subsequent log records and applies them in LSN sequence to generate a new page version (step 4), which is then added to the Page Store buffer pool for future access (step 5). The buffer pool functions as a write-back cache, asynchronously flushing dirty pages to the slice log (step 6) allowing it to apply multiple log records before writing a page to disk, further reducing the amount of I/O. Once a new page version is flushed, the Log Directory is updated to point to the new location. Log records and page versions that are older than the recycle LSN reported by the SAL can be safely removed. 

The Page Store buffer pool serves as a second-level cache for the buffer pools of the database front end. However, its primary function is to reduce disk reads during consolidation rather than help with foreground reads. We have evaluated both LFU and LRU policies for the Page Store buffer pool and found that LFU provides a 25\% better hit rate. This is consistent with previous research showing that an LFU policy is better suited for second-level caches \cite{Li05}. 

There is a separate Log Directory for each slice. This reduces contention and the size of the key. On the other hand, the log cache and Buffer pool are global to the Page Store to exploit differences in slice activity automatically. Most slices have little update activity, but some slices are write-intensive and need a larger share of memory for incoming log records. 

The algorithm for selecting pages to consolidate is critical to the performance of the Page Store as it ultimately determines both the I/O required and the memory consumed by the Log Directory and the log cache. Our initial choice was a "the longest chain first" approach. To minimize page writes, we cycled through all slices and chose the pages with the longest chains of log records for consolidation. However, this policy gives priority to hot pages and leaves out relatively cold pages with few log records. With many unconsolidated cold pages, the number of the log records becomes excessive, and they get evicted from the cache in FIFO order. With time, the number of evicted unconsolidated records becomes large, resulting in a large memory footprint for the Log Directory and generating many small read requests to bring log records back into memory. This makes consolidation even slower.

To avoid a flood of small I/O operations, we adopted a consolidation policy that is "log cache-centric". Pages to be consolidated are chosen in the order that their log records arrive in the Page Store and appear in the log cache. If the log cache is full, new log fragments are saved to disk and added to a queue of log fragments to be loaded into the log cache as soon as space becomes available. Only log records that are in the log cache are used to produce new versions of pages. As soon as a log record has been consolidated, it is removed from the log cache. This ensures that consolidation tasks operate only on in-memory log records and the data is never read from disk. This "log cache-centric" approach results in a lower buffer pool hit rate but completely eliminates log cache misses. The benefits of steady and predictable performance outweigh the lower buffer pool hit rate.

\section{Experimental evaluation}
In this section, we evaluate the performance of Taurus and compare it with competing solutions. As a baseline for comparison, we have chosen Amazon Aurora as one of the most popular DBaaS implementations, the recently introduced Microsoft Socrates, and the community version of MySQL 8.0 running with locally-attached storage.

\subsection{Comparison with Amazon Aurora}\label{comparison-with-aurora-section}
For the comparison with Aurora, we used the same benchmarks and hardware similar to those used by Verbitski et al. \cite{Verbitski17}. We ran the MySQL front end on a machine with 32 vCPUs and 256GB of memory, 80\% of which was dedicated to the database buffer pool. We used newer Xeon 6161 CPUs instead of the E5-2670 CPU used by Aurora, but we ran at lower clock frequency (2.2GHz vs. 2.5GHz) to compensate for possible architectural improvements. Aurora and Taurus both use modified versions of MySQL as the front end.

\begin{figure}[t]
\begin{center}
\includegraphics[width=0.9\linewidth]{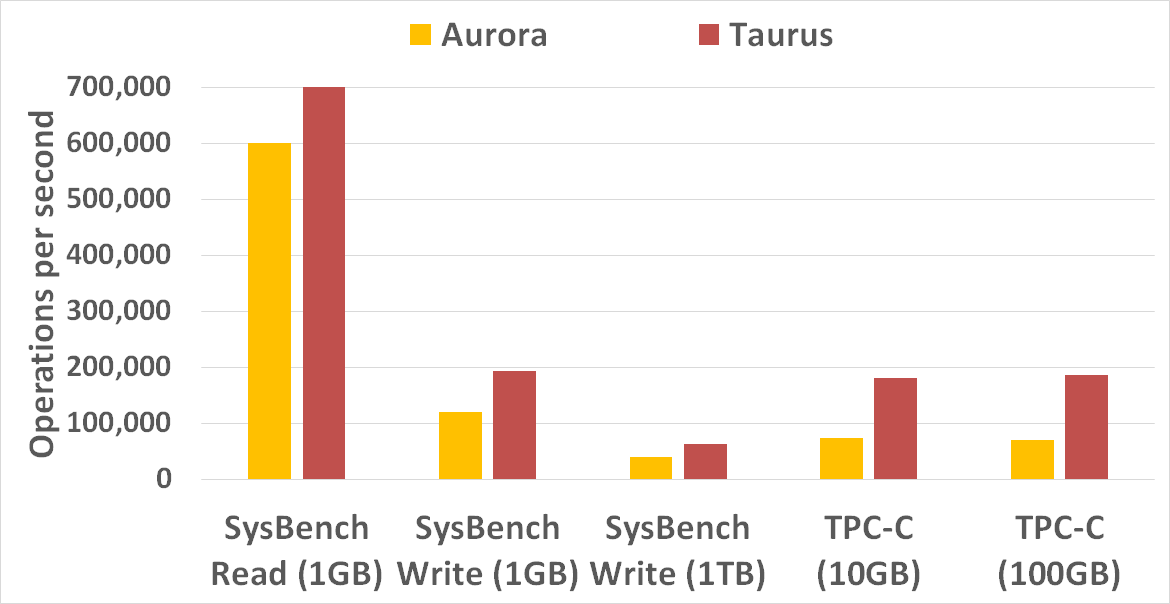}
\end{center}
\caption{Taurus performance in benchmarks
\label{results-vs-aurora}}
\end{figure}

We ran the SysBench read-only and write-only workloads as well as the Percona TPC-C variant  \cite{PerconaTPC} with databases of different sizes. Fig.~\ref{results-vs-aurora} shows the results. The vertical axis represents the number of reads per second for the SysBench read-only, writes per second for the SysBench write-only, and transactions per minute (tpmC) for the TPC-C benchmarks. Taurus outperforms the results published by the Aurora team in all five benchmarks. In the read-only test the difference is small (16\%), but in the write-only benchmark, the Taurus advantage is more than 50\% - and reaches 160\%  in TPC-C. Without knowing the  Aurora internals, we can only speculate what causes the performance improvement. We believe that the replacement of quorum-based replication with the novel Taurus replication strategy is a major contributing factor.

\subsection{Comparison with Microsoft Socrates}
While Taurus and Aurora both use MySQL as the database front end, Socrates uses SQL Server. This and the lack of details of the hardware environment makes comparisons more difficult. The Socrates team compared their performance with the performance of SQL Server running against locally-attached storage on a similar hardware configuration. We compare performance results for Socrates relative to SQL Server reported in \cite{Antonopoulos19} with the performance of Taurus relative to MySQL 8.0, also running with local storage of the same type as the storage used in Taurus storage layer.

\begin{figure}[t]
\begin{center}
\includegraphics[width=0.9\linewidth]{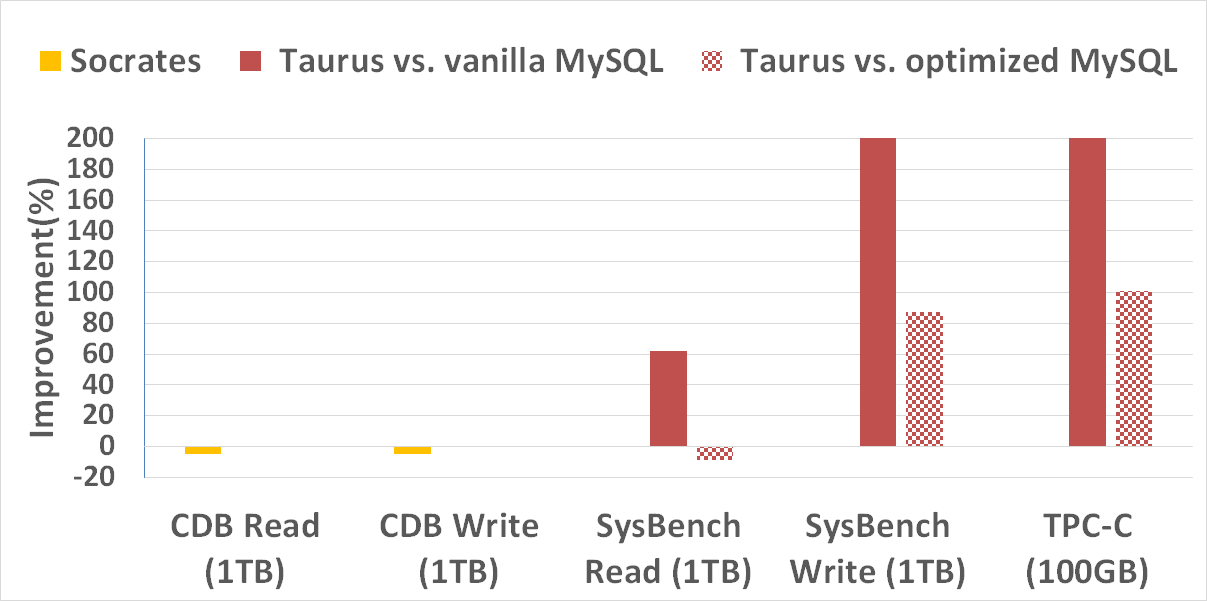}
\end{center}
\caption{Performance relative to a monolithic database with local storage
\label{results-vs-socrates}}
\end{figure}

Fig.~\ref{results-vs-socrates} displays the results. The first two experiments correspond to the published Socrates results \cite{Antonopoulos19}. Socrates achieves performance slightly (5\%) worse than SQL Server. In comparison, Taurus demonstrates improved performance relative to vanilla MySQL with local storage ranging from 50\% for the SysBench read-only workload (third solid column) to a 200\% improvement for the SysBench write-only workload and TPC-C (fourth and fifth solid columns). We believe that Taurus performs better than MySQL with local storage due to Taurus's fast network hardware and append-only writes, while MySQL relies on write-in-place, which is slower. While Socrates also used state of the art network hardware, Taurus compares better against a local-storage configuration. We believe that this is because Taurus has just two network-separated tiers, while Socrates requires four.

Both Aurora and Socrates introduced incremental optimizations to the database front end. In Taurus, we have also done similar optimizations. To separate the performance impact of the Taurus architecture innovations from the impact of the front end optimizations, we ported the front end optimizations to the local-storage version of MySQL. The results are presented as cross-hatched columns in the last three experiments in Fig.~\ref{results-vs-socrates}. On the read-only workload, Taurus is 9\% slower than the optimized version of MySQL with locally attached storage due to the higher network latency of the remote storage. On the write-only and TPC-C workloads, our architectural changes allow Taurus to outperform the optimized MySQL  by 87\% and 101\%, respectively.


\subsection{Read replicas}
\begin{figure}[t]
\begin{center}
\includegraphics[width=0.9\linewidth]{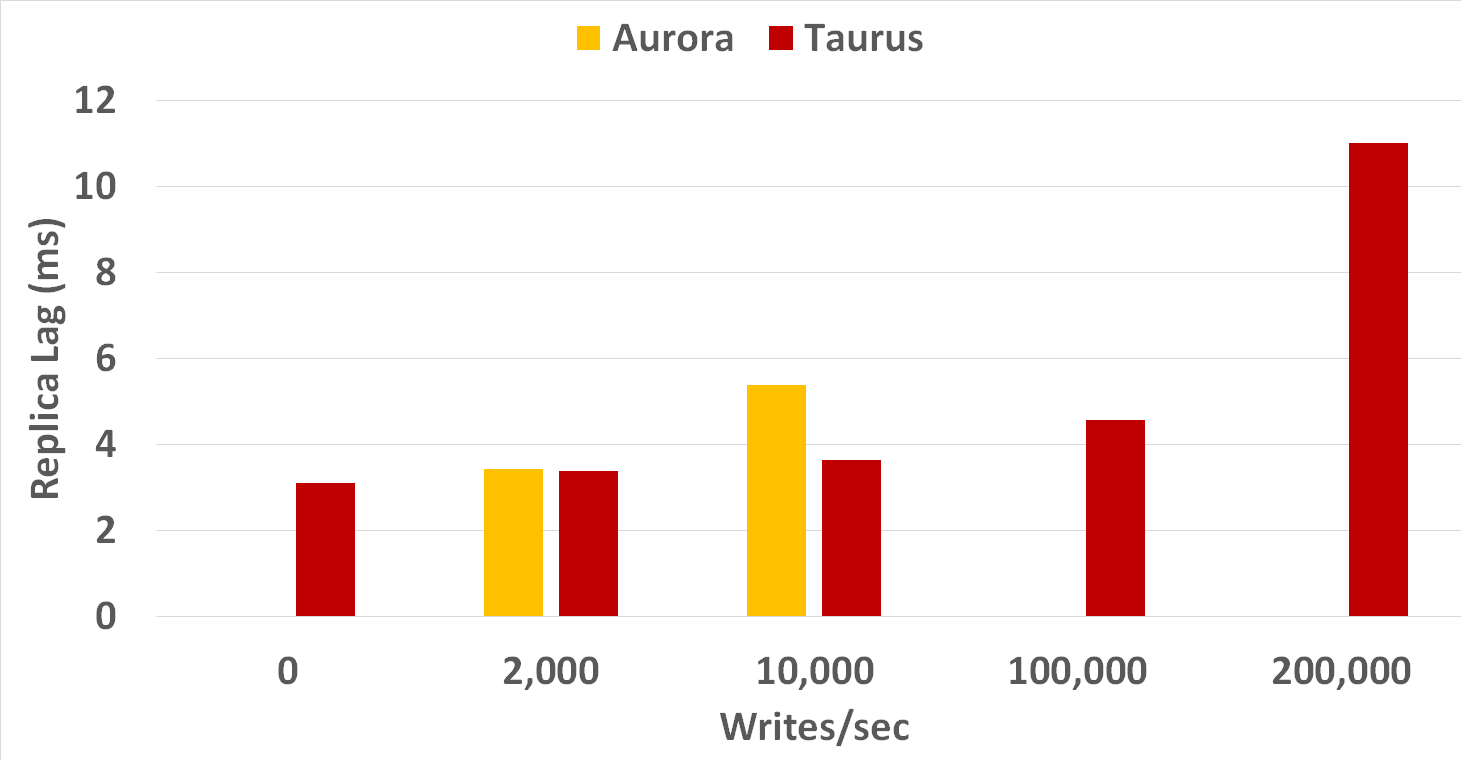}
\end{center}
\caption{Replica Lag for SysBench Write-Only workload
\label{results-replica-lag}}
\end{figure}

One of the most important characteristics of multi-replica solutions is the lag between the master and read replicas. In order to measure this lag, we run a  SysBench write-only workload on the master, varying the number of connections in order to achieve the desired level of writes per second. The replica is not running any read-only workload. We measure a time difference between the time a value is changed by the master and the time when the new value is read by the replica. To avoid additional latency of client-server communications, we use stored procedures to update and check values on the master and the replica. We collected multiple measurements and averaged the result. The results are presented in Fig.~\ref{results-replica-lag}. A comparison with corresponding available Aurora results  published by the Aurora team \cite{Verbitski17} is provided for reference. For a light load on the master, Taurus replicas show low lag values, similar to or lower than those demonstrated by Aurora.  In addition, Taurus demonstrates good workload scalability even when the master executes heavy workloads up to 200,000 writes per second. At this utilization level, the replica lag is below 11ms, which is small enough for many applications. The Aurora authors do not provide data above 10,000 writes per second. This low replica lag, even when the master is very busy, is due to the Taurus design decision not to send logs directly to read replicas from master, but let read replicas to read them from Log Stores. The network interface of the master, is no longer a bottleneck, and network traffic is distributed across multiple Log Stores. This also allows us to support a large number of read replicas without the master becoming a bottleneck.

\section{Conclusion}
This paper presents Taurus, a new cloud-native relational database. Taurus is based on a "the log is the database" paradigm and separates compute and storage layers. This paper describes the implementation choices and optimizations that enabled the improved performance observed in our experiments.

The Taurus architecture and new replication algorithm result in availability higher than that of traditional quorum-based replication without sacrificing performance or hardware costs. The  replication algorithm is based on separate persistence mechanisms for database logs and for pages. It combines strong and eventual consistency models to optimize performance and availability.

Future work includes moving more tasks from the compute layer to the more scalable storage layer, supporting multi-master capability, and using the latest advancements in hardware including storage-class memory and RDMA.

\begin{acks}
Many people provided invaluable feedback and advice on the initial design of the Taurus database, including Hao Feng, Robin Grosman, Tianzheng Wang, Xun Xue, Lei Zhang, Qingqing Zhou. This work would not also be possible without people who contributed to the Taurus development, including Zhaorong Chen, Gengyuan Dan, Ge Dang, Yu Du, Bo Gao, Kareem El Gebaly, Jianwei Li, Liyong Song, Hongbin Lu, Juncai Meng, Yuanyuan Nie, Samiao Ren, Weiyi Ruan, Calvin Sun, Guojun Wu, Lengdong Wu, Andrew Xiong, Peng Xu, Cheng Zhao, Rui Zhou, and all members of the Huawei product development team.  We also want to express our gratitude to the anonymous reviewers and Gord Sissons for their valuable comments.
 
\end{acks}

\newpage
\bibliographystyle{ACM-Reference-Format}
\bibliography{taurus-bib}

\newpage
\appendix
\begin{figure}[H]
\begin{center}
\includegraphics[width=0.9\linewidth]{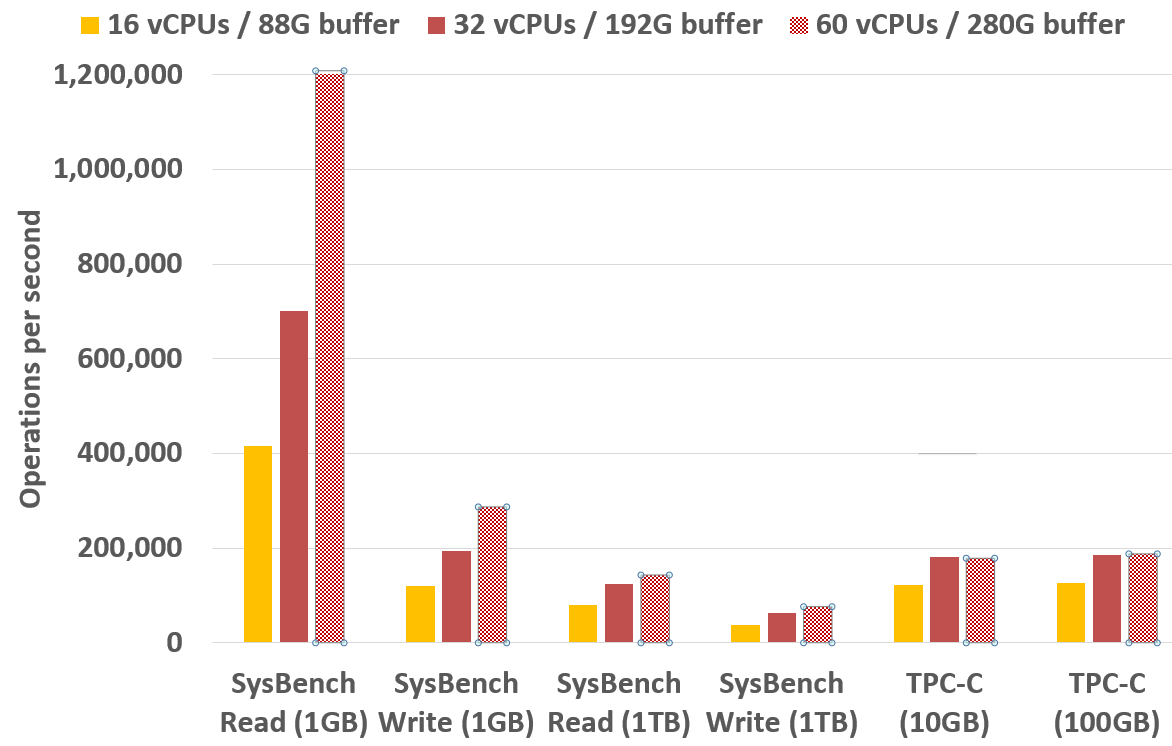}
\end{center}
\caption{Scaling with instance size
\label{results-scalability}}
\end{figure}

\begin{figure}[H]
\begin{center}
\includegraphics[width=0.9\linewidth]{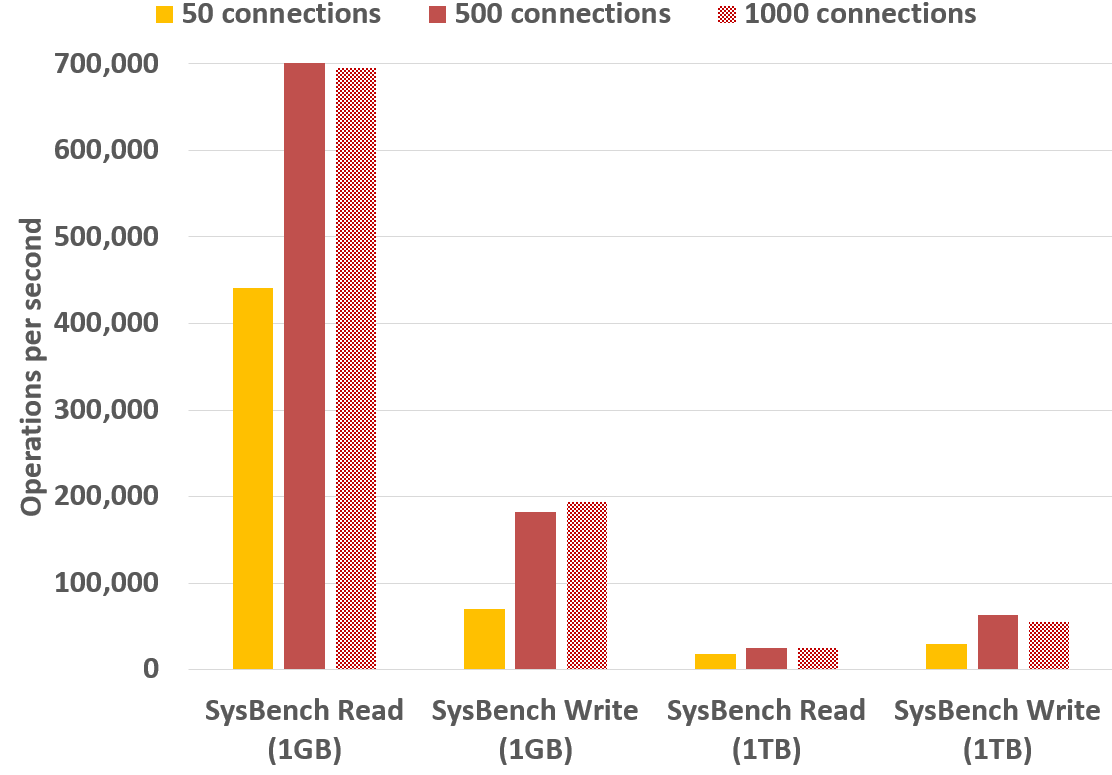}
\end{center}
\caption{Scaling with number of connections
\label{results-connections}}
\end{figure}

\section{Additional experimental evaluation}

In this section, we provide experimental results that further describe Taurus performance and scalability.

\subsection{Scaling with database front end node sizes}
In this experiment, we measure the ability of Taurus to scale up with the capacity of the database front end node. The results are presented in Fig.~\ref{results-scalability} for three different node instances. The three instances considered have 16, 32, and 60 vCPUs. They have the database front end buffer pool size of 88GB, 192GB, and 280GB respectively. SysBench read and write workloads for 1GB database scales linearly with node capacity. With a 1TB database, SysBench performance also grows with node capacity, although the growth is sub-linear. TPC-C benchmark performance does not change much between 32 and 60 vCPUs nodes due to database data contention.

\begin{figure}[H]
\begin{center}
\includegraphics[width=0.9\linewidth]{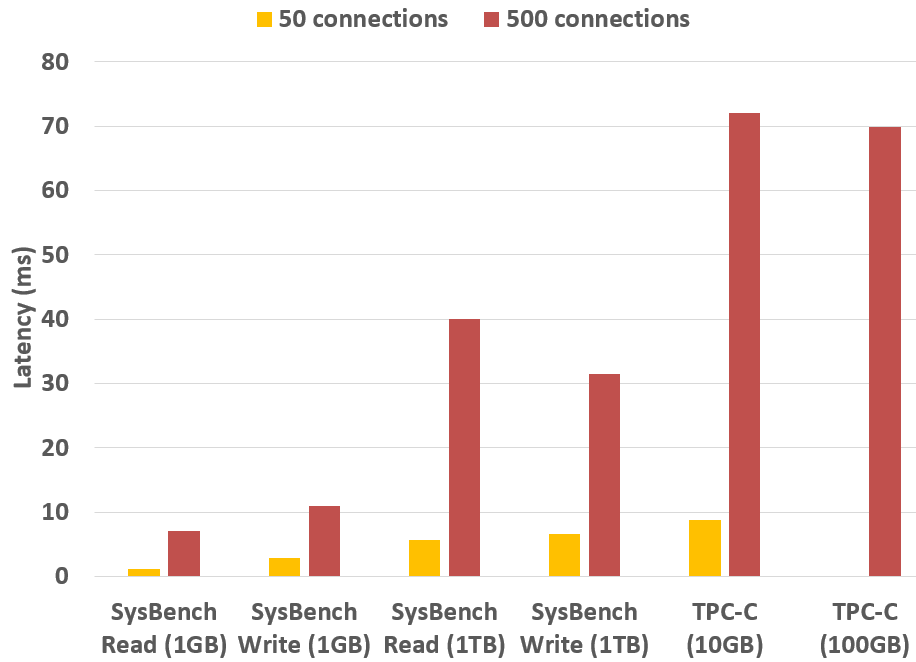}
\end{center}
\caption{Query latency
\label{results-latency}}
\end{figure}

\subsection{Scaling with number of connections}
In this experiment, we use the same database front end node with 32 vCPUs and 192GB buffer pool as in the previous experiment and vary the number of simultaneous client connections. The results are displayed on Fig.~\ref{results-connections}. Taurus scales up to 500 connections and further increasing the number of connections to 1000 does not result in increased performance.

\subsection{Query latency}
In this experiment, we measure query latency for SysBench and TPC-C benchmarks. We use the same experimental setup as described in the "\ref{comparison-with-aurora-section} Comparison with Amazon Aurora subsection". The results are presented in Fig.~\ref{results-latency}. The SysBench Read benchmark for 1GB and 1TB databases is particularly interesting as it shows the upper bound of overhead of separating  compute and storage layers. With a 1GB database, all page reads are done from the database front end buffer pool and resulting in latency slightly more than 1ms for 50 simultaneous connections. With the 1TB database, most of requests need to go to the storage layer and latency increases to 5 ms. This additional latency includes network communication, accessing log directories, and reading pages from storage devices.

\end{document}